
%
\input phyzzx
\PHYSREV
\vsize=54pc
\normalspace
\unnumberedchapters
\date={March 1993}
\Pubnum={\caps UPR-547-T
 }

\def\to{\rightarrow}
\titlepage
\title{ Determination of $Z'$ Gauge
Couplings to Quarks and Leptons at Future Hadron Colliders
}
\frontpageskip=0.5\medskipamount plus 0.5 fil
\author{ F. del Aguila$^+$, Mirjam Cveti\v c,$^* $and  Paul Langacker$^*$}
\address{$^*$Department of Physics\break
University of Pennsylvania\break
Philadelphia, PA 19104--6396\break
 and\break
$^+$Department of Physics\break
University of Granada\break
Granada, Spain\break}

 \abstract{
We point out that  at future hadron colliders
the ratio of cross sections for $pp\to Z'\to \ell^+\ell^-$  in
two  rapidity bins  is a useful probe of
the relative couplings of the $Z'$ to
$u$ and $d$ quarks. Combined  with the forward-backward
asymmetry, the rare decay modes  $Z'\to W \ell\nu_\ell$,
and three associated  productions $pp\to Z'V$ ($V=Z,W,\gamma$),
and assuming inter-family universality,  small $Z-Z'$ mixing,
and  the $Z'$ charge commuting with the $SU_{2L}$ generators,  three
out of four normalized couplings
could be extracted. An analysis of the statistical
uncertainties expected for the above  probes
at  the LHC  for  typical models
with $M_{Z'}\simeq1$ TeV shows that one  lepton coupling and
two combinations of
quark couplings could be determined to around 5\%,
20\%, and 30\%, respectively. This allows for a clear distinction between
models. }

\line{PACS \# 12.15, 12.10, 11.15\hfill}
\endpage

 \REF\LRR{ P.~Langacker, R.~Robinett, and J.~Rosner, Phys. Rev.
D {\bf30}, 1470 (1984).}
 \REF\Bar{ V.~Barger, {\it et al.}, Phys. Rev.   {\bf D 35}, 2893
(1987).}
\REF\DuLa{ L.~Durkin and P.~Langacker, Phys. Lett. {\bf B166},
436 (1986); F.~del~Aguila, M.~Quiros, and F.~Zwirner, Nucl. Phys. B
{\bf287}, 419 (1987); {\bf284}, 530 (1987); J.~Hewett and T.~Rizzo in
{\sl Proceedings of the 1988 Snowmass Summer Study on High Energy
Physics in the 1990's}, Snowmass, CO 1988; P.~Chiappetta  {\it et
al.}, in the {\sl Proceedings of the Large Hadron Collider Workshop},
Aachen, Germany, 1990; J. Hewett and T. Rizzo, Phys. Rep. {\bf 183}, 193
(1989).}
\REF\HRII{J. ~Hewett and  T. ~Rizzo, MAD/PH/649/91.}
 A heavy gauge boson
$Z'$ could be produced
 and clearly detected via leptonic decays   $pp \to Z'\to \ell^+
\ell^-$ ($\ell=e,\mu$)
at the LHC and SSC if its mass does not exceed around
5 TeV\rlap.\refmark{\LRR -\HRII}\
 The immediate goal after the discovery of a new gauge boson would be to
understand its origin and properties, including
its couplings to ordinary fermions,
the  nature of the symmetry breaking, and its couplings to exotic fermions
and supersymmetric partners.

\REF\CLII{M. Cveti\v c and P. Langacker,
 Phys. Rev. {\bf D 46}, R14 (1992).}
\REF\CKL{M.~Cveti\v c, B.~Kayser, and P.~Langacker,
Phys. Rev. Lett. {\bf68}, 2871 (1992).}
\REF\PACO{F. del Aguila, B. Alles, Ll. Ametller and
A. Grau, University of Granada preprint, UG-FT-22/92/Rev (December 1992).}
\REF\HR{J. Hewett and T. Rizzo, Argonne National Laboratory
preprint, ANL-HEP-PR-92-33 (June 1992).}
\REF\CLIV{M. Cveti\v c and
P. Langacker,
Phys. Rev. {\bf D46}, 4943 (1992).}
\REF\RII{T. Rizzo, Phys. Rev. {\bf D47}, 965 (1993).}
Recently there has been a renewed
interest in diagnostic probes of the couplings of possible
heavy $Z'$ gauge bosons to ordinary
fermions at future hadron colliders.\refmark{\CLII -\RII }
The forward-backward
asymmetry\refmark\LRR\
in the main  production channel  $pp\to Z'\to
\ell^+\ell^- \ (\ell=e$ or $\mu$) has long been known
to be useful.\Ref\DAV{See also F. del Aguila and J. Vidal,
Int. Journal of Math. Phys. {\bf A4}, 4097 (1989).}
It is now understood, however, that several
complementary
probes\refmark{\CLII ,\CLIV ,\RII} would be useful
for $M_{Z'} < (1 -2)\ TeV$.

\REF\RI{T. Rizzo, Phys. Lett. B {\bf 192}, 125 (1987).}

 In particular,
rare decays\refmark{\RI, \CLII}  were recognized
and studied in detail\refmark{\CLII, \PACO , \HR}.
Such decays involve $Z'\rightarrow f_1\overline{f_2} V$,
with  ordinary  gauge bosons $V=(Z,W)$ bremsstrahled from one of the
fermionic  ($f_{1,2}$) legs.
A background study\refmark{\CLII, \CLIV} of
such  decays
revealed that the only useful mode without
large standard model and QCD backgrounds
is $Z'\to W\ell \nu_{\ell}$  and $W\to hadrons$,
 with the imposed  cut $m_{T\ell\nu_\ell}>90$ GeV on the transverse mass of the
$\ell \nu_\ell$.
(This assumes that there is a sufficiently high efficiency for the
  reconstruction of $W\to hadrons$ in events tagged by an energetic lepton.)
 The same mode   with
$W\to \ell\nu_\ell$ may also   be detectable\refmark\PACO\  if appropriate
cuts are applied.
 These modes probe a particular combination of $Z'$ gauge couplings to
leptons.

 Associated productions
$pp\rightarrow Z' V$
with $V=(Z,W)$  and $Z'\to \ell^+\ell^-$
were recently proposed\refmark\CLIV\ to
probe the gauge couplings to
 quarks, and are thus complementary to  rare decays.
\Ref\OTHER{The $\tau$ polarization
in $pp\to Z'\to \tau^+\tau
^-$ would be another useful probe if it can be measured. See
J. Anderson, M. Austern, and B. Cahn, Phys. Rev. {\bf D46}, 290 (1992), Phys.
Rev. Lett. {\bf 69}, 25 (1992).
Similarly, if  proton polarization were available the measurements of the
corresponding asymmetries in $pp\to Z'\to \ell
^+\ell ^-$ would also be useful. See A. Fiandrino and  P. Taxil, Phys. Lett.
{\bf B292}, 242 (1992) and references therein.}
The associated $Z'$ production
with $V=\gamma$ was also proposed.\refmark\RII\

Rare decays and
associated production involve  processes with four-fermion
final states, and thus have suppressed rates  compared
to the main production channels $pp\to Z'\to \ell^+  \ell^- $.

 In this paper we  point out that due to the
 harder valence $u$-quark distribution in the proton
 relative to the $d$-quark,
the ratio $r_{y_1}$ of production
cross-sections in the
two rapidity bins ($ |y| =\{0,y_1\}$ and $|y| =\{y_1,y_{max}\}$)
is a useful complementary probe for separating the $Z'$ couplings to
the $u$ and $d$ quarks. 
 We choose $y_1$  in such a way
that each bin has
a comparable number of events in order to minimize the
 statistical error.

Another purpose of this paper is to examine how well the various $Z'$
couplings could be extracted from the  above six  signals
 at  future colliders.
 For definiteness, we consider the statistical uncertainties for a
1 TeV  $M_{Z'}$ at the LHC with a projected luminosity of
 $10^{34} \hbox{cm}^{-2}\hbox{s}^{-1} $. At the SSC with
 $10^{33} \hbox{cm}^{-2}\hbox{s}^{-1} $ one expects about
 half as many events.
Eventually,
the uncertainties associated with the detector acceptances and  systematic
errors will have to be taken into account.

{\it Formalism and Typical Models}.
\REF\LL{P. Langacker and M. Luo, Phys. Rev. {\bf D 44}, 817 (1991).}
The neutral current gauge interaction term in the presence of an
additional $U_1$ is
$-L_{NC}=eJ_{em}^\mu A_\mu +g_1 J_1^\mu Z_{1\mu}+ g_2J_2^\mu Z_{2\mu},
$
with $Z_1$ being the $SU_2 \times U_1$ boson and $Z_2$ the additional
boson in the weak eigenstate basis.  Here
$g_1\equiv\sqrt{g_L^2+g_Y^2}=g/\cos\theta_W$, where
 $g_L$, $g_Y$ are the
gauge couplings of $SU_{2L}$ and $U_{1Y}$,
and $g_2$ is the gauge coupling of $Z_2$.
The currents   are:
$J_j^\mu={1\over2}\sum_i{\bar\psi_i}\gamma^\mu
\left[\hat{g}^i_{V_j}-\hat{g}^i_{A_j}\gamma_5\right] \psi_i,
\quad j=1,2,
$
where the sum runs over fermions, and the $\hat{g}^i_{(V,A)_j}$
are the vector and axial vector couplings of $Z_j$ to the
$i^{th}$ flavor.  Analogously, $\hat{g}^i_{(L,R)_j}={1\over2}(
\hat{g}^i_{V_j}\pm \hat{g}^i_{A_j})$.

We consider the following typical
 GUT, left-right  symmetric, and
superstring-motivated models:
{ \it{(i)}}$Z_\chi$ occurs in $SO_{10}\rightarrow SU_5\times U_{1\chi}$,
{\it{(ii)}}$Z_\psi$ occurs in $E_6\rightarrow SO_{10}\times U_{1\psi}$,
 {\it{(iii)}}$Z_\eta=\sqrt{3/8}Z_\chi-\sqrt{5/8}Z_\psi$ occurs in
superstring inspired models in which $E_6$ breaks directly to a rank 5
group,
{\it{(iv)}}$Z_{LR}$ occurs in left-right  (LR) symmetric models. Here
we consider the special value
$\kappa=g_R/g_L=1 $ of the gauge couplings $g_{L,R}$ for $SU_{2L,2R}$,
respectively.

In the rest of the paper we assume family universality
 and neglect $Z-Z'$
mixing (as  suggested from  experiments).
We also assume $[Q',T_i]=0$,
where $Q'$ is the $Z'$ charge  and $T_i$ are the $SU_{2L}$ generators,
which holds for $SU_2 \times U_1 \times U_1'$ and LR models.
The relevant quantities\refmark\CLIV\  to
 distinguish different theories are
the charges, $\hat{g}^u_{L2}=\hat{g}^d_{L2}\equiv\hat{g}^q_{L2}$,
$\hat{g}^u_{R2}$, $\hat{g}^d_{R2}$, $\hat{g}^\nu_{L2}=\hat{g}^e_{L2}
\equiv\hat{g}^\ell_{L2}$, and $\hat{g}^\ell_{R2}$, and the gauge
coupling strength $g_2$.  The overall scale of the charges (and $g_2$)
depends on the normalization convention for $\Tr(Q'^2)$,
but the ratios characterize particular theories.
The signs of the charges will be hard to determine at hadron
colliders. Some information is possible in principle from $\gamma$
and $Z$ interference effects, but this is expected to be small.
Other possibilities include
precision experiments
and possible future $e^+e^-$ colliders.
We therefore concentrate only on the four ``normalized''
observables:\refmark\CLIV
\def\denom{{(\hat{g}^\ell_{L2})^2+(\hat{g}^\ell_{R2})^2}}
$$\gamma_L^\ell\equiv{{(\hat{g}^\ell_{L2})^2}\over\denom},\ \
\gamma_L^q\equiv{{(\hat{g}^q_{L2})^2}\over\denom},\ \
\tilde{U}\equiv\left({{\hat{g}^u_{R2}}\over {\hat{g}^q_{L2}}}\right)^2,\ \
\tilde{D}\equiv\left({{\hat{g}^d_{R2}}\over
{\hat{g}^q_{L2}}}\right)^2.\eqn\tild$$
The values of  $\gamma_L^\ell$, $\gamma_L^q$,
 $\tilde{U}$, and $\tilde{D}$  for models {\it (i)-(iv)} are listed in
Table~I.

 {\it  Rapidity Ratio.} In the main production channels
$pp\to Z'\to\ell^+\ell^-$ ($\ell=e,\mu$)
we define the ratio:
$$r_{y_1}= {{\int_{-y_1}^{y_1}{{d\sigma}\over {dy}}dy}\over
{(\int_{-y_{max}}^{-y_{1}}+\int_{y_1}^{y_{max}}) {{d\sigma}\over
{dy}}dy}}={{\int_{-y_1}^{y_1}[F(y)+B(y)]dy}
\over{(\int_{-y_{max}}^{-y_{1}}+\int_{y_1}^{y_{max}}) [F(y)+B(y)]dy}},
\eqn\ry$$
 where $F(y)\pm B(y)=[\int_0^1\pm\int_{-1}^0] \,d\cos\theta (d^2\sigma/
dy\, d\cos\theta)$ and $\theta$ is the $\ell^-$ angle in the $Z'$ rest
frame. The rapidity range is  from $\{ -y_{max}, y_{max}\}$.
 $y_1$ is chosen in a range $0<y_1<y_{max}$ so that the
number of events in the two bins are comparable.
 At the LHC $y_{max}\simeq 2.8$ for  $M_{Z'}\simeq1$ TeV,  and $y_1=1$ turns
out to be an
appropriate choice.

$r_{y_1}$ can be expressed
 in terms of $\tilde U$ and $\tilde D$. The expression for
 $M_{Z'}=1$ TeV at the LHC is given in the
first line of the Table II, using the quark distributions of
Ref.~\Ref\EHLQ{E. Eichten, I.
Hinchliffe, K. Lane and C. Quigg, Rev. Mod. Phys. {\bf 56}, 579 (1984).}\ .
 This expression and those for the
other probes are adequate for illustration. The use of other
structure functions leads to somewhat different expressions. If
a $Z'$ is actually observed it would be necessary to recalculate the
expressions using updated distribution functions (which should by then be known
to a few \%), and QCD corrections to the $Z'$ production would have to be
included.

The numerator and denominator involve different combinations of
$\tilde U$ and $\tilde D$, reflecting the harder distribution of valence
$u$ quarks. In particular, the dependence on $\tilde U$ and
$\tilde D$ is sufficiently different from that of the
forward-backward asymmetry $A_{FB}$ (see the
second
 line\Ref\FOOTXV{
In Ref.\refmark\CLIV\ the coefficient $0.38\equiv{3\over 4}\times 0.51$
was erroneously quoted to be ${3\over 4}\times 0.58$.} of
 Table II),\refmark\CLIV\  that $r_{y_1}$ provides a  complementary probe.
For the typical models described here  the values for
$r_{y_1}$ and
their statistical  errors\Ref\FOOTIII{The statistical errors are
based on the same (illustrative) branching ratio assumptions
as in Ref.\refmark{\CLII} .}
(for $Z'\to [e^+e^-+\mu^+\mu^-]$)
are given for $M_{Z'}=1$ TeV
at the  LHC  in the first line of Table II.
The statistical errors are  sufficiently small for
$r_{y_1}$ to be  useful for distinguishing between the models.

Another potential possibility is the
ratio of forward-backward asymmetries in the two
rapidity bins. We define:
$$A_{FB_{y1}}={{(\int_{0}^{y_1}-\int_{-y_1}^{0}) [F(y)-B(y)]dy}
\over{(\int_{y_1}^{y_{max}}-\int_{-y_{max}}^{-y_{1}})
[F(y)-B(y)]dy}},
\eqn\AFBY$$
where $F(y),\ B(y)$ are defined after Eq.\ry\ .
$A_{FB_{y1}}$ can be viewed as ``a refinement of a refinement''
in the main production channel, since it involves
 the angular distribution of $\ell^\pm$  as well as  the rapidity
distribution.\Ref\FOOT{
The quantities $r_{y_1}$, $A_{FB}$, and $A_{FB_{y1}}$ are useful for
displaying the dependences of the production distribution on the
underlying couplings. In
practice, however, it would be more efficient to directly fit to the observed
distribution in
$y$ and $\theta$, folding in the detector  acceptances and other
systematic uncertainties.}
The expression for $A_{FB_{y1}}$  is given in the third line of Table II.
{}From the expression and the explicit numerical values
for typical models (see Table II) it is apparent that $A_{FB_{y1}}$ is
{ not} a sensitive
enough function of the gauge couplings to provide useful information for the
projected luminosities.

For completeness in Table II  we also quote results for  the rare decay
mode  and the associated productions. For the ``gold-plated''
events\Ref\FOOTXXXI{The
 only other signal which does not suffer from a major standard model or QCD
background is
$Z'\to Z\ell^+\ell^-$.  This, however, turns out {\it not} to be a useful
diagnostic probe. See Refs. \refmark\CLII\ and \refmark\CLIV\ for details.}
$Z'\to W\ell\nu_{\ell}$  and $W\to hadrons$  the ratio\Ref\FOOTXII{This ratio
is
independent of
 the branching ratio $B(Z'\to \ell^+\ell^-)$ and therefore
probes the gauge couplings of the ordinary fermions only.}
$r_{\ell\nu W} \equiv {{B(Z'\rightarrow W\ell\nu_\ell)}
\over {B(Z'\rightarrow\ell^+\ell^-)}}$  is defined, in which one sums
over $\ell=e,\mu$ and  over $W^+$, $W^-$.
  $r_{\ell\nu W}$ is rewritten  in terms of the gauge
couplings
in the the fourth line of Table
II,\Ref\FOOXXIII{ The coefficient is
$0.067\equiv 0.77\ a_W$,
where  $a_{W}\simeq{\alpha\over{6\pi \cos\theta_W^2\sin\theta
_W^2}}
\left[ \ln^2\mu +3\ln\mu
+5 -{\pi^2\over3}\right]$
is a  kinematic factor which only depends on
 $\mu=M^2_{W}/M^2_{Z'}$.
  For $\alpha=1/128$ and $M_{Z'}=1$ TeV, $a_W$ = $0.087$. In
Ref.\refmark\CLIV\  $\alpha=1/137$ and thus  $a_W=0.080$ were used.
}
along with the values  and statistical error bars for  the
typical  models.
It is apparent that this decay  is an excellent probe of
$\gamma_L^\ell$.

For the associated productions
 one   defines\refmark\CLIV\
 the ratios:
$R_{Z'V}={{\sigma
(pp\to Z'V)B(Z'\to \ell^+\ell^-)}\over{
\sigma (pp\to Z')B(Z'\to \ell^+\ell^-)}}$
 with $V=(Z, W)$\ \refmark\CLIV\ and $V=\gamma$\ \refmark\RII\ decaying into
leptons and  quarks, and $\ell$ includes both
$e$ and $\mu$.
The expressions\Ref\FOOTXX{ In
 Ref.\refmark\CLIV\ slightly different numerical values
 for $R_{Z'Z}$ and $R_{Z'W}$  were given, mainly due to less accurate
numerical integrations.} and values for these ratios are given  in Table II
for $M_{Z'}=1 $ TeV at the LHC. (For
$R_{Z'\gamma} $ a transverse momentum cut $p_{T\gamma}>50$ GeV is imposed.)
The ratios $R$ yield direct
 information on the couplings of {\it  quarks} to
$Z'$.   $R_{Z'(Z,W)}$ primarily single out a  combination
 $\sim 2\tilde{U}+\tilde{D}$,
which is the same quantity
that is probed by $A_{FB}$ (for a known $\gamma_L^\ell$), but the extra
information provides a welcome consistency check.
The numerator in the expression for $R_{Z'Z}$
has a weak dependence  on  $\tilde{U}$ and $\tilde{D}$,
due to the fact that they are weighted
by the squares of the (small)
gauge couplings of the right-handed quarks to the $Z$.
 $R_{Z'\gamma}$  has a strong dependence on $\tilde U$ in the numerator.
For the typical models (except LR)  $\tilde U=1$, and thus
$R_{Z'\gamma}$ by itself is not a useful discriminant between these models.

Within the  assumptions of
interfamily universality, negligible $Z-Z'$ mixing, and $[Q',T_i]$ $= 0$
 one sees from Table II that  the six quantities
$r_{y_1}$, $A_{FB}$, $r_{\ell\nu W}$, and $R_{Z'V}$ with $(V=Z,W,\gamma)$
yield significant information on  three ($\gamma_L^\ell$, $\tilde U$ and
$\tilde
D$) out of four
 normalized gauge couplings\Ref\FOOTIA{
 The overall normalization is not fixed by the ratios. It could be determined
 by an independent measurement of the $Z'$ width.}
 of ordinary fermions to the $Z'$.
The relative size of the $Z'$ couplings to quarks and leptons
could be determined by a measurement of the branching ratio
 $B(Z'\rightarrow q{\bar q})$. In particular,  the ratio ${1\over3}
{{B(Z'\rightarrow q{\bar q})}\over{B(Z'\rightarrow\ell^+\ell^-)}}=
   \gamma_L^q (2+\tilde U+\tilde D)$
(counting all 3 families)
 would yield
  the left-handed quark coupling\refmark\CLIV\   $\gamma_L^q$. However,
this appears difficult.\Ref\RM{A. Henriques and L. Poggioli, ATLAS
Collaboration, Note PHYS-NO-010 (October 1992);
T. Rizzo, ANL-HEP-PR-93-18 (March 1993).
See also P. Mohapatra, Univ. of Colorado preprint 92-0580 (June 1992).}

{\it Determination of the Couplings.} To study  with
what precision these couplings could be determined, we have performed
a  combined $\chi^2$ analysis of the observables $r_{y_1}$, $A_{FB}$,
  $r_{\ell\nu W}$, and $R_{Z'V}$ with $V=(Z,W,\gamma)$ for each of the models.
We have included only the statistical uncertainties (from Table II), and have
ignored correlations between the observations.
\Ref\FOOTXXX {
In  principle there is a correlation between the errors of the
input quantities $ A_{FB}$ and
$r_{y_1}$, since they both depend on four quantities
$F_1\equiv \int_0^{y_1} F(y)dy +\int_{-y_1}^0 B(y)dy$,
$F_2\equiv \int_{y_{1}}^{y_{max}} F(y)dy + \int_{-y_{max}}^{-y_{1}} B(y)dy$,
and similar definitions for $B_{1,2}$.
However, explicit calculation yields the correlation  $\rho= \left[
\sqrt{r_{y_1}}/(1+r_{y_1}) \right]
\times (A_{FB_1}-A_{FB_2})/\sqrt{1-A_{FB}^2}$.
Namely, it  is
proportional to a  difference of forward-backward asymmetries
 $A_{FB_{1,2}}\equiv  (F_{1,2}-B_{1,2})/(F_{1,2}+B_{1,2})$
in the two rapidity  bins.
 $A_{FB_{1,2}}$ and their differences turn out to be
numerically small, and therefore the correlation is negligible.}
The resulting uncertainties for
the couplings are  given
in  Table I.

In particular, $\gamma_L^\ell$ can be determined very well (between
2\% and 8\% for the $\chi$, $\psi$, and $\eta$ models), primarily due to
the small statistical error for the rare decay mode $Z'\to W\ell\nu_\ell$.
On the other hand the  quark couplings have larger uncertainties,
typically 20\% for $\tilde U$, and an absolute error of $\sim 0.3 - 0.6$ for
$\tilde D$ (except $Z_{LR}$).

{}From the explicit dependence of the
probes on $\gamma_L^\ell$ , $\tilde U$ and $\tilde D$ (see Table II)
one sees that the
correlation between $\tilde U$ and $\tilde D$ is appreciable, while
$\gamma_L^\ell$ is weakly correlated with $\tilde U$ and $ \tilde D$ because of
the small statistical error on $r_{\ell\nu W}$, which singles out
$\gamma_L^\ell$. Explicit calculation shows that this is the case for all the
models studied except for $Z_\chi$.  In this case the statistical errors on
$r_{\ell\nu W}$ and $A_{FB}$ (which depends on all
three variables) are comparable, inducing
sizable correlations. The fitted correlation coefficient  between
$\tilde U$ and $\tilde D$ is given  for each model
in the last line of  Table I.

In Figs. 1a, 1b and 1c the 1 $\sigma$ ($\Delta\chi^2=1$) and 90\% confidence
level  ($\Delta \chi^2=4.6$) contours are plotted
for $\tilde D$ versus $\gamma_L^\ell$, $\tilde U$ versus $\gamma_L^\ell$, and
$\tilde D$ versus $\tilde U$, respectively. The statistical error bars are for
 $M_{Z'}=1$ TeV
 at the LHC for the  $\eta$, $\psi$ and $\chi$ models. The $LR$ model
has  $\tilde U$ and $\tilde D$ in different region of parameter space  (see
Table I).
 From Figures 1a and 1b it is
clear  that  one can distinguish
well between  different  models. In Figure 1c
the correlations between $\tilde U$ and $\tilde D$  are
evident, while  from  Figures 1a and 1b   the correlation between
$\gamma_L^\ell$ and ($\tilde U , \tilde D$)  is significant only for the
 $Z_\chi$.

{\it Conclusions.}
In this note we have explored  possible experimental signals which  probe
 $Z'$ gauge couplings  to ordinary fermions at hadron colliders. In addition to
the  forward-backward asymmetry\refmark\LRR\  $A_{FB}$, and the
 more recently
 proposed  rare decay modes  $Z'\to W\ell\nu_\ell$\refmark\CLII\ and
associated productions $pp\to Z'V$ with $V=Z,W$\refmark\CLIV\ and
$V=\gamma$\refmark\RII , we point out that the ratio $r_{y_1}$ of the
 cross-sections in the two rapidity bins  in the main production channels
 is a useful
complementary  probe of the relative $Z'$ couplings to $u$ and $d$ quarks.

To test the sensitivity of the six
 proposed signals
we express them in terms of the normalized gauge couplings to
quarks and leptons; as an example we chose $M_{Z'}=1$ TeV at the LHC.
The error analysis  shows that under the assumptions of family universality,
small $Z - Z'$ mixing, and
$[Q',T_i]=0$ the  magnitude of three out of four gauge couplings could be
determined with precisions of around 5\% for $\gamma_L^\ell$,  around 20\% for
$\tilde U$, and somewhat higher for  $\tilde D$, allowing  for the clear
identification of  particular models.

For higher $Z'$ masses the number of events drops rapidly. For
$M_{Z'}=2$ TeV, the statistical errors on $r_{y1}$, $A_{FB}$,
and $r_{\ell \nu W}$ increase by a factor of  4, while those on
$R_{Z'V}$ increase by a factor of 3. From Tables I and II we see that
reasonable discrimination between models and determination of
the normalized parameters is still possible. However, for
$M_{Z'}=3$ TeV the statistical errors on the first three quantities
are by a factor of 13 larger than for 1 TeV, and there are not enough
events expected for $R_{Z'V}$ to allow a meaningful measurement.
 For $M_{Z'}\geq 3$ TeV, there is therefore little ability
to discriminate  between models.

This work was supported by the Department of Energy Grant \#
DE--AC02--76--ERO--3071, and an SSC fellowship award (M.C.).
\filbreak
\eject
%
\newbox\hdbox%
\newcount\hdrows%
\newcount\multispancount%
\newcount\ncase%
\newcount\ncols
\newcount\nrows%
\newcount\nspan%
\newcount\ntemp%
\newdimen\hdsize%
\newdimen\newhdsize%
\newdimen\parasize%
\newdimen\spreadwidth%
\newdimen\thicksize%
\newdimen\thinsize%
\newdimen\tablewidth%
\newif\ifcentertables%
\newif\ifendsize%
\newif\iffirstrow%
\newif\iftableinfo%
\newtoks\dbt%
\newtoks\hdtks%
\newtoks\savetks%
\newtoks\tableLETtokens%
\newtoks\tabletokens%
\newtoks\widthspec%
%
%
\immediate\write15{%
CP SMSG GJMSINK TEXTABLE --> TABLE MACROS V. 851121 JOB = \jobname%
}%
%
%
\tableinfotrue%
\catcode`\@=11
%
%
\def\tstrut{\vrule height3.1ex depth1.2ex width0pt}%
\def\and{\char`\&}
\def\tablerule{\noalign{\hrule height\thinsize depth0pt}}%
\thicksize=1.5pt
\thinsize=0.6pt
\def\thickrule{\noalign{\hrule height\thicksize depth0pt}}%
\def\ctr#1{\hfil\ #1\hfil}%
%
%
%
%
\tablewidth=-\maxdimen%
\spreadwidth=-\maxdimen%
\def\tabskipglue{0pt plus 1fil minus 1fil}%
%
%
\centertablestrue%
%
%
%
%
\parasize=4in%
\gdef\ARGS{########}
\gdef\headerARGS{####}
\def\@mpersand{&}
{\catcode`\|=13
\gdef\letbarzero{\let|0}
\gdef\letbartab{\def|{&&}}%
\gdef\letvbbar{\let\vb|}%
}
{\catcode`\&=4
\def\ampskip{&\omit\hfil&}
\catcode`\&=13
\let&0
\xdef\letampskip{\def&{\ampskip}}%
\gdef\letnovbamp{\let\novb&\let\tab&}
}
\def\begintable{
   \begingroup%
   \catcode`\|=13\letbartab\letvbbar%
   \catcode`\&=13\letampskip\letnovbamp%
   \def\multispan##1{
      \omit \mscount##1%
      \multiply\mscount\tw@\advance\mscount\m@ne%
      \loop\ifnum\mscount>\@ne \sp@n\repeat%
   }
   \def\|{%
      &\omit\widevline&%
   }%
   \ruledtable
}
\long\def\ruledtable#1\endtable{%
%
%
%
   \offinterlineskip
   \tabskip 0pt
   \def\widevline{\vrule width\thicksize}
   \def\endrow{\@mpersand\omit\hfil\crnorm\@mpersand}%
   \def\crthick{\@mpersand\crnorm\thickrule\@mpersand}%
   \def\crthickneg##1{\@mpersand\crnorm\thickrule
          \noalign{{\skip0=##1\vskip-\skip0}}\@mpersand}%
   \def\crnorule{\@mpersand\crnorm\@mpersand}%
   \def\crnoruleneg##1{\@mpersand\crnorm
          \noalign{{\skip0=##1\vskip-\skip0}}\@mpersand}%
   \let\nr=\crnorule
   \def\endtable{\@mpersand\crnorm\thickrule}%
   \let\crnorm=\cr
%
%
   \edef\cr{\@mpersand\crnorm\tablerule\@mpersand}%
   \def\crneg##1{\@mpersand\crnorm\tablerule
          \noalign{{\skip0=##1\vskip-\skip0}}\@mpersand}%
   \let\ctneg=\crthickneg
   \let\nrneg=\crnoruleneg
   \the\tableLETtokens
%
%
   \tabletokens={&#1}
%
%
   \countROWS\tabletokens\into\nrows%
   \countCOLS\tabletokens\into\ncols%
%
%
   \advance\ncols by -1%
   \divide\ncols by 2%
   \advance\nrows by 1%
%
%
   \iftableinfo %
      \immediate\write16{[Nrows=\the\nrows, Ncols=\the\ncols]}%
   \fi%
%
%
   \ifcentertables
      \ifhmode \par\fi
      \line{
      \hss
   \else %
      \hbox{%
   \fi
      \vbox{%
         \makePREAMBLE{\the\ncols}
         \edef\next{\preamble}
         \let\preamble=\next
         \makeTABLE{\preamble}{\tabletokens}
      }
      \ifcentertables \hss}\else }\fi
   \endgroup
   \tablewidth=-\maxdimen
   \spreadwidth=-\maxdimen
}
\def\makeTABLE#1#2{
   {
   \let\ifmath0
   \let\header0
   \let\multispan0
%
%
   \ncase=0%
   \ifdim\tablewidth>-\maxdimen \ncase=1\fi%
   \ifdim\spreadwidth>-\maxdimen \ncase=2\fi%
   \relax
%
   \ifcase\ncase %
      \widthspec={}%
   \or %
      \widthspec=\expandafter{\expandafter t\expandafter o%
                 \the\tablewidth}%
   \else %
      \widthspec=\expandafter{\expandafter s\expandafter p\expandafter r%
                 \expandafter e\expandafter a\expandafter d%
                 \the\spreadwidth}%
   \fi %
   \xdef\next{
      \halign\the\widthspec{%
      #1
      \noalign{\hrule height\thicksize depth0pt}
      \the#2\endtable
%
      }
   }
   }
   \next
}
\def\makePREAMBLE#1{
   \ncols=#1
   \begingroup
   \let\ARGS=0
   \edef\xtp{\widevline\ARGS\tabskip\tabskipglue%
   &\ctr{\ARGS}\tstrut}
   \advance\ncols by -1
   \loop
      \ifnum\ncols>0 %
      \advance\ncols by -1%
      \edef\xtp{\xtp&\vrule width\thinsize\ARGS&\ctr{\ARGS}}%
   \repeat
   \xdef\preamble{\xtp&\widevline\ARGS\tabskip0pt%
   \crnorm}
   \endgroup
}
\def\countROWS#1\into#2{
   \let\countREGISTER=#2%
   \countREGISTER=0%
   \expandafter\ROWcount\the#1\endcount%
}%
\def\ROWcount{%
   \afterassignment\subROWcount\let\next= %
}%
\def\subROWcount{%
   \ifx\next\endcount %
      \let\next=\relax%
   \else%
      \ncase=0%
      \ifx\next\cr %
         \global\advance\countREGISTER by 1%
         \ncase=0%
      \fi%
      \ifx\next\endrow %
         \global\advance\countREGISTER by 1%
         \ncase=0%
      \fi%
      \ifx\next\crthick %
         \global\advance\countREGISTER by 1%
         \ncase=0%
      \fi%
      \ifx\next\crnorule %
         \global\advance\countREGISTER by 1%
         \ncase=0%
      \fi%
      \ifx\next\crthickneg %
         \global\advance\countREGISTER by 1%
         \ncase=0%
      \fi%
      \ifx\next\crnoruleneg %
         \global\advance\countREGISTER by 1%
         \ncase=0%
      \fi%
      \ifx\next\crneg %
         \global\advance\countREGISTER by 1%
         \ncase=0%
      \fi%
      \ifx\next\header %
         \ncase=1%
      \fi%
      \relax%
      \ifcase\ncase %
         \let\next\ROWcount%
      \or %
         \let\next\argROWskip%
      \else %
      \fi%
   \fi%
   \next%
}
\def\counthdROWS#1\into#2{%
\dvr{10}%
   \let\countREGISTER=#2%
   \countREGISTER=0%
\dvr{11}%
\dvr{13}%
   \expandafter\hdROWcount\the#1\endcount%
\dvr{12}%
}%
\def\hdROWcount{%
   \afterassignment\subhdROWcount\let\next= %
}%
\def\subhdROWcount{%
   \ifx\next\endcount %
      \let\next=\relax%
   \else%
      \ncase=0%
      \ifx\next\cr %
         \global\advance\countREGISTER by 1%
         \ncase=0%
      \fi%
      \ifx\next\endrow %
         \global\advance\countREGISTER by 1%
         \ncase=0%
      \fi%
      \ifx\next\crthick %
         \global\advance\countREGISTER by 1%
         \ncase=0%
      \fi%
      \ifx\next\crnorule %
         \global\advance\countREGISTER by 1%
         \ncase=0%
      \fi%
      \ifx\next\header %
         \ncase=1%
      \fi%
\relax%
      \ifcase\ncase %
         \let\next\hdROWcount%
      \or%
         \let\next\arghdROWskip%
      \else %
      \fi%
   \fi%
   \next%
}%
{\catcode`\|=13\letbartab
\gdef\countCOLS#1\into#2{%
   \let\countREGISTER=#2%
   \global\countREGISTER=0%
   \global\multispancount=0%
   \global\firstrowtrue
   \expandafter\COLcount\the#1\endcount%
   \global\advance\countREGISTER by 3%
   \global\advance\countREGISTER by -\multispancount
}%
\gdef\COLcount{%
   \afterassignment\subCOLcount\let\next= %
}%
{\catcode`\&=13%
\gdef\subCOLcount{%
   \ifx\next\endcount %
      \let\next=\relax%
   \else%
      \ncase=0%
      \iffirstrow
         \ifx\next& %
            \global\advance\countREGISTER by 2%
            \ncase=0%
         \fi%
         \ifx\next\span %
            \global\advance\countREGISTER by 1%
            \ncase=0%
         \fi%
         \ifx\next| %
            \global\advance\countREGISTER by 2%
            \ncase=0%
         \fi
         \ifx\next\|
            \global\advance\countREGISTER by 2%
            \ncase=0%
         \fi
         \ifx\next\multispan
            \ncase=1%
            \global\advance\multispancount by 1%
         \fi
         \ifx\next\header
            \ncase=2%
         \fi
         \ifx\next\cr       \global\firstrowfalse \fi
         \ifx\next\endrow   \global\firstrowfalse \fi
         \ifx\next\crthick  \global\firstrowfalse \fi
         \ifx\next\crnorule \global\firstrowfalse \fi
         \ifx\next\crnoruleneg \global\firstrowfalse \fi
         \ifx\next\crthickneg  \global\firstrowfalse \fi
         \ifx\next\crneg       \global\firstrowfalse \fi
      \fi
\relax
      \ifcase\ncase %
         \let\next\COLcount%
      \or %
         \let\next\spancount%
      \or %
         \let\next\argCOLskip%
      \else %
      \fi %
   \fi%
   \next%
}%
\gdef\argROWskip#1{%
   \let\next\ROWcount \next%
}
\gdef\arghdROWskip#1{%
   \let\next\ROWcount \next%
}
\gdef\argCOLskip#1{%
   \let\next\COLcount \next%
}
}
}
\def\spancount#1{
   \nspan=#1\multiply\nspan by 2\advance\nspan by -1%
   \global\advance \countREGISTER by \nspan
   \let\next\COLcount \next}%
\def\dvr#1{\relax}%
\def\header#1{%
\dvr{1}{\let\cr=\@mpersand%
\hdtks={#1}%
\counthdROWS\hdtks\into\hdrows%
\advance\hdrows by 1%
\ifnum\hdrows=0 \hdrows=1 \fi%
\dvr{5}\makehdPREAMBLE{\the\hdrows}%
\dvr{6}\getHDdimen{#1}%
{\parindent=0pt\hsize=\hdsize{\let\ifmath0%
\xdef\next{\valign{\headerpreamble #1\crnorm}}}\dvr{7}\next\dvr{8}%
}%
}\dvr{2}}
\def\makehdPREAMBLE#1{
\dvr{3}%
\hdrows=#1
{
\let\headerARGS=0%
\let\cr=\crnorm%
\edef\xtp{\vfil\hfil\hbox{\headerARGS}\hfil\vfil}%
\advance\hdrows by -1
\loop
\ifnum\hdrows>0%
\advance\hdrows by -1%
\edef\xtp{\xtp&\vfil\hfil\hbox{\headerARGS}\hfil\vfil}%
\repeat%
\xdef\headerpreamble{\xtp\crcr}%
}
\dvr{4}}
\def\getHDdimen#1{%
\hdsize=0pt%
\getsize#1\cr\end\cr%
}
\def\getsize#1\cr{%
\endsizefalse\savetks={#1}%
\expandafter\lookend\the\savetks\cr%
\relax \ifendsize \let\next\relax \else%
\setbox\hdbox=\hbox{#1}\newhdsize=1.0\wd\hdbox%
\ifdim\newhdsize>\hdsize \hdsize=\newhdsize \fi%
\let\next\getsize \fi%
\next%
}%
\def\lookend{\afterassignment\sublookend\let\looknext= }%
\def\sublookend{\relax%
\ifx\looknext\cr %
\let\looknext\relax \else %
   \relax
   \ifx\looknext\end \global\endsizetrue \fi%
   \let\looknext=\lookend%
    \fi \looknext%
}%
%
%
\def\tablelet#1{%
   \tableLETtokens=\expandafter{\the\tableLETtokens #1}%
}%
\catcode`\@=12
\noindent{Table I}
\vskip 0.2in
\begintable
\| $\chi$\ \ \  & $\psi$\ \ \  & $\eta$\ \ \ & $LR$\ \ \ \crthick
$\gamma^\ell_L$ \| $0.9\pm 0.018$ & $0.5\pm 0.03$& $0.2\pm 0.015$ &
$0.36\pm 0.007$
\nr
$\gamma^q_L$ \| 0.1\ \ \  & 0.5\ \ \  & 0.8\ \ \  & 0.04\ \ \ \nr
$\tilde{U}$ \| $1\pm 0.18$ & $1\pm 0.27$ & $1\pm 0.14$ & $37\pm 8.3$ \nr
$\tilde{D}$ \| $9\pm 0.61$ & $1\pm 0.41$ & $0.25\pm 0.29$ & $65\pm 14$ \nr
$\rho_{ud}$\| -0.19& -0.24& -0.66&0.93\endtable
\vskip0.2in
\noindent{\caps Table~I} Values of $\gamma_L^\ell$, $\gamma_L^q$,
 $\tilde{U}$,
and $\tilde{D}$
for the $\chi$, $\psi$, $\eta$, and ${LR}$
 models. The error bars indicate how well the coupling could be
measured at the  LHC  for $M_{Z'}=1$ TeV. $\rho_{ud}$ indicates the correlation
coefficient between $\tilde U$ and $\tilde D$.
Except for the $\chi$ model the correlation  between $\gamma_L^\ell$
and ($\tilde U, \  \tilde D$) are negligible.

\vskip 0.2in
\noindent{Table II}
\vskip 0.2in
 \begintable
   & $\chi$ & $\psi$ & $\eta$ & ${LR}$
\crthick
$r_{y_1}
=1.55{{1+0.64 \tilde U+ 0.36 \tilde D}\over{1+0.73\tilde
U+0.27\tilde D}}$
& $1.79\pm 0.02$ & $1.55\pm 0.04$ &
  $1.49\pm 0.03$ & $1.62\pm 0.014$  \nr
$A_{FB}
=0.38\left(2\gamma_L^\ell-1\right) {{1
-0.75\tilde{U}-0.25\tilde{D}}\over{1
+0.68\tilde{U}+0.32\tilde{D}}}$ & $-0.134\pm0.007$ & $0.000\pm0.016$ &
  $-0.025\pm0.014$ & $0.098\pm0.006$  \nr
$A_{FB_{y1}}
=0.60{{1-0.73\tilde U-.27\tilde D}
\over{1-0.76\tilde U-0.24\tilde D}}
$ & $0.68\pm0.08$ & ---  &
  $0.68\pm0.88$ & $0.61\pm.08$  \crthick
$r_{\ell\nu W}=0.067\gamma_L^\ell$ &$   0.060\pm 0.0014
$& $ 0.034\pm 0.002
$&$ 0.013\pm 0.001
$&
$  0.024\pm 0.0008
$\nr
$R_{Z'Z}=10^{-3}{{7.94+0.96\tilde{U}+0.11\tilde{D}}\over
{1+0.68\tilde{U}+0.32\tilde{D}}}$ &  $\ 0.0022\pm 0.0002$
& $
 0.0045\pm 0.0008$& $   0.0051\pm 0.0007
 $& $   0.0011\pm 0.0001$\nr
$R_{Z'W}=10^{-3}{{25.7}\over
{1+0.68\tilde{U}+0.32
\tilde{D}}}$ &$   0.0056\pm 0.0004
$& $   0.013\pm 0.001$&$  0.015\pm 0.001
$&
$   0.00055\pm 0.00010$\nr
$R_{Z'\gamma}=10^{-3}5.62{{1+0.89\tilde U+0.11\tilde D}\over
{1+0.68\tilde{U}+0.32
\tilde{D}}}$ &$ 0.0035\pm 0.0003
$& $ 0.0056\pm 0.0009$&$ 0.0061\pm 0.0008
$&
$ 0.0049\pm 0.0003$ \endtable

 Table II. The quantities
$r_{y_1}$, $A_{FB}$,  $A_{FB_{y1}}$,
  $r_{\ell\nu W}$, and $R_{Z'V}$ with $V=(Z,W,\gamma)$
and their
numerical values  (with
statistical  errors)   for $M_{Z'}=1$ TeV
at the LHC.  Error bars for $r_{y_1}$, $r_{\ell\nu W}$, $R_{Z'V}$ are for
$e + \mu$ channels, while $A_{FB}$ and $A_{FB_{y1}}$ are for $e$
{\it or} $\mu$.
\endpage
{\bf Figure caption}

Figure 1. 1 $\sigma$ ($\Delta \chi^2 = 1$) contours (solid lines) and
 the 90\% confidence level ($\Delta \chi^2 = 4.6$) contours (dotted lines)
 for the $\chi$, $\psi$ and
$\eta$ models are plotted for $\tilde D$ versus $\gamma_L^\ell$ (Figure 1a),
$\tilde U$ versus $\gamma_L^\ell$ (Figure 1b), and $\tilde D$ versus $\tilde U$
(Figure
1c). The input data are for $M_{Z'}=1$ TeV  at the LHC
and include statistical errors only.

\endpage

\refout\end